\documentclass[doublecol]{epl2} 

\usepackage{amsmath}
\usepackage{command}

\title{Pumping viscoelastic two-fluid media}
\shorttitle{Pumping viscoelastic two-fluid media} 

\author{H. Wada}
\shortauthor{H. Wada}

\institute{Yukawa Institute for Theoretical Physics, Kyoto University, Kyoto 606-8502, Japan                   
}
\pacs{87.16.Uv}{Active transport process}
\pacs{47.15.G-}{Low-Reynolds-number (creeping) flows}
\pacs{83.60.Bc}{Linear viscoelasticity}

\abstract{
Using a two-fluid model for viscoelastic polymer solutions, we study analytically fluid transport driven by a transverse, small
amplitude traveling wave propagation.
The pumping flow far from the waving boundary is shown to be strongly wave number and 
viscosity dependent, in contrast to a viscous Newtonian fluid.
We find the two qualitatively different regimes:
In one regime relevant to small wave numbers, the fluidic transport is almost the same as the Newtonian case, and uniform viscoelastic constitutive
equations provide a good approximation. In the other regime, the pumping is substantially decreased
because of the gel-like character. The boundary separating these two regimes is clarified.
Our results suggest possible needs of two-fluid descriptions for the transport and locomotion in biological fluids with cilia and flagella.}

\begin{document}

\maketitle

\section{Introduction}
Active fluid transport or mixing in a complex fluid finds many examples
on scales of cellular biology~\cite{Lighthill-Review-1976,Brennen-Review-1977,Childress-Book}.
In the fluid transport in respiratory or digestive organs~\cite{Blake-Review-1988}, 
large arrays of beating cilia on a surface organize metachronal waves, in which neighboring cilia beat in a constant phase lag, 
to pump mucus fluids~\cite{Gueron-PNAS-1997,Kim-Netz-PRL-2006,Stark-EPL-2009}.
During development of an embryo, flows generated by arrays of rotating nodal cilia dictate the left-right symmetry breaking 
in the placement of organs~\cite{Nonaka-Nature-2002}.
In plant cells, cytoplasmic streaming, which is responsible for rapid intracellular transport and mixing, is driven by the motor protein myosin 
that moves along filamentary actin at the cell periphery~\cite{Goldstein-PRL-2008}.

Hydrodynamic thrust generated by surface distortions is also used to propel many prokaryotic and eukaryotic cells~\cite{Stone-PRL-1996}.
A sperm cell beats its single flagellum and swims through mucus in the female mammalian reproductive tract~\cite{Fauci-Review-2006}.
Microorganisms such as {\it Paramecium} or {\it Volvox} colonies possess ciliated surfaces, and propel themselves with metachronal
coordination.
As an example of locomotion in contact with a substrate, amoeboid cells crawl accompanying the protoplasmic streaming 
driven by periodic contractile cell body deformations~\cite{Nakagaki-BJ-2008}. 

All these examples involve small-scale, slow flows to which the zero Reynolds number hydrodynamics is applied.
In addition, fluid media can also be viscoelastic, such as a gel-forming character of mucus and protoplasmic sol in amoeboid cells. 
While the basic principles of swimming at zero Reynolds numbers in a Newtonian
fluid have been established for years~\cite{Taylor-Royal-1951, Lighthill-Review-1976, Brennen-Review-1977,Childress-Book, Lauga-Powers-2009,Ishikawa-Interface-2009}, 
the role of viscoelasticity on flagellar motility or transport is still elusive.

The importance of nonlinear effects of a fluid elasticity has been addressed in recent studies. 
Employing various constitutive equations with fading memories, Lauga investigated 
a sheet with a given wave pattern and showed that the fluid elasticity always
reduces its swimming speed~\cite{Lauga-PhysFluids-2007,Lauga-EPL-2009}. 
Fu {\it et al.} extended the findings of Lauga to a beating filament, and discussed 
the changes in the beating pattern and of the swimming direction~\cite{Fu-PRL-2007}.
Teran {\it et al}. studied the problem numerically and demonstrated the enhancement of the swimming ability for large 
tail undulations~\cite{Teran-PRL-2010}. 
All these works employed uniform constitutive equations for non-Newtonian fluids such as Stokes-Oldroyd-B models.
However, in constrast to a Newtonian fluid acting as a uniform viscous background, viscoelastic media, such as polymer 
solutions and gels, are essentially dynamically strucutred materials with characteristic time and length scales. 
For example, transient network structures are speculated to be responsible for a mysterious viscosity-dependent swimming of 
some spiral bacteria~\cite{Berg-Turner-1979,Kaiser-Nature-1975,Nakamura-BJ-2006, Wada-Netz-PRE-2009,Leshansky-PRE-2009}.  
To our knowledge, no theoretical attempt has been made to combine a mesoscopic description for complex fluids and
a theory of active fluid transport and locomotion.

\begin{figure}
\begin{center}
 \onefigure[width=0.99\linewidth]{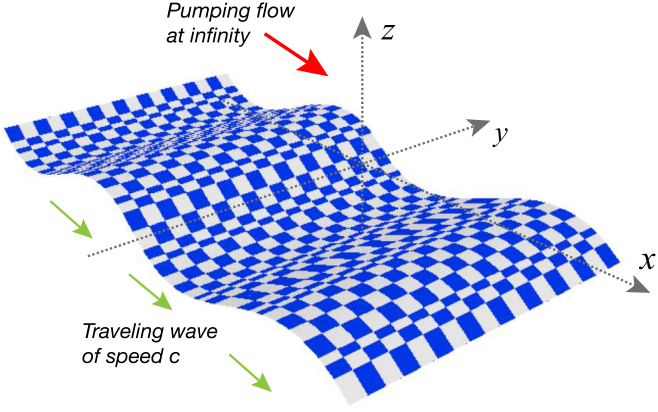}
\caption{Schematics of a waving sheet immersed in a viscoelastic fluid.
The sheet (boundary) is infinitely extended and is fixed in the laboratory frame (except its undulating motion).
The system is assumed to be uniform along $y$-direction.
A surface disturbance is a traveling wave of phase speed $c$ and a net pumping flow is generated at inifinity.}
\label{fig:schematics}
\end{center}
\end{figure}

In this letter, employing a two-fluid description for viscoelastic fluids~\cite{Tanaka-JCP-1973,deGennes-Macromol-1977,Doi-Onuki-1992,Levine-PRE-2001}, 
we consider the two-dimensional hydrodynamic flow field driven by a planar flexible boundary with a transverse propagating wave
of small amplitude~\cite{Taylor-Royal-1951}, 
and discuss the induced pumping flow at infinity (see fig.~\ref{fig:schematics}).
This net flow corresponds to a swimming speed in the lab frame when the sheet is allowed to move 
translationally~\cite{Taylor-Royal-1951,Stone-PRL-1996}.
We show that the hydrodynamic pumping flow generated far from the waving boundary is strongly wave number and viscosity dependent.
For large wave numbers, the pumping transport is substantially decreased compared to the Newtonian case, and 
its asymptotic value is determined by the elastic properties of the medium.
A mesoscale fluid property is crucial. The diffusive relaxation of the viscoelastic network stress and its frictional 
coupling to the viscous Newtonian solvent effectively acts as a sink of the velocity field, leading to the decrease of the far field pumping flow.
Our analysis is thus complementary to the previous studies~\cite{Lauga-PhysFluids-2007,Fu-PRL-2007,Lauga-EPL-2009}, 
where spatially uniform viscoelastic stress relaxations were postulated.

\section{Model}
In general, the hydrodynamic flow profile at the far field is rather insensitive to its near-filed details.
Therefore, many essential features and symmetries of flagellum propulsion or pumping with planar and helical waves may be captured by
the present minimal model. 
To proceed, we give a profile of the flexible boundary as a traveling wave of amplitude $b$
\begin{eqnarray}
 z(x,t) &=& \eps b\sin(kx-\omega t),
 \label{eq:wave-pattern}
\end{eqnarray}
where we have introduced the small dimensionless expansion parameter $\eps$, which we will set unity eventually. 
The wave number $k$ is taken positive without loss of generality.
The displacement, eq.~(\ref{eq:wave-pattern}), is independent of a load, and the surrounding fluid satisfies the no-slip dounary conditions on the deforming sheet surface. 

We describe our model viscoelastic medium as a polymer network with mesh size $\xi_0$ coupled to a viscous Newtonian solvent
with viscosity $\eta$ via a friction constant $\Gamma$, estimated as $\Gamma\simeq \eta/\xi_0^2$~\cite{
Tanaka-JCP-1973, deGennes-Macromol-1977, Doi-Onuki-1992,Levine-PRE-2001}.
The polymer network is described as a continuum elastic medium with shear and bulk Lame coefficients $\mu$ and $\lambda$.
Neglecting all inertial contributions unimportant to our aim, the linearized equations of motion for the network displacement
${\bf u}$ and the fluid velocity ${\bf v}$ are~\cite{Levine-PRE-2001,Powers-PRE-2008} 
\begin{eqnarray}
 \mu\nabla^2{\bf u}+(\lambda+\mu)\nabla(\nabla\cdot{\bf u}) &=& \Gamma(\dot{\bf u}-{\bf v}),
 \label{eq:eqm-01}\\
 -\nabla p+\eta\nabla^2{\bf v} &=& -\Gamma(\dot{\bf u}-{\bf v}),
 \label{eq:eqm-02}\\
 \nabla\cdot[(1-\phi){\bf v}+\phi\dot{\bf u}] &=& 0,
 \label{eq:eqm-03}
\end{eqnarray}
where the dot represents the time derivative, and the incompressibility of the whole fluid is required in eq.~(\ref{eq:eqm-03}).
Since the volume fraction of the polymer $\phi$ is sufficiently low even for entangled solutions, eq.~(\ref{eq:eqm-03}) is
simplified as $\nabla\cdot{\bf v}=0$~\cite{Levine-PRE-2001,Powers-PRE-2008}.
We identify the elastic stress tensor of the network as $\vecsg^p=\mu[\nabla{\bf u}+(\nabla{\bf u})^T]+\lambda\nabla(\nabla\cdot{\bf u})$ 
and define a viscoelastic force ${\bf f}^p=\nabla\cdot\vecsg^p=\Gamma({\bf v}-\dot{\bf u})$. 
For an entangled polymer solution, the elastic network is actually formed only transiently; 
the elastic stress can be relaxed through disentanglements of polymer chains~\cite{Doi-Onuki-1992,Onuki-Book}.
This effect is most easily incorporated by adding a term $\tau^{-1} {\bf f}^p$ to the dynamical equation for ${\bf f}^p$ 
derived from eq.~(\ref{eq:eqm-01})-(\ref{eq:eqm-03}): 
\begin{equation}
  \dot{\bf f}^p = -\frac{1}{\tau}{\bf f}^p+
  	\mu\nabla^2{\bf v}+\frac{\mu}{\Gamma}\nabla^2{\bf f}^p+\frac{\mu+\lambda}{\Gamma}\nabla(\nabla\cdot{\bf f}^p),
 \label{eq:eqm-04}
\end{equation}
where $\tau$ is the characteristic stress relaxation time~\cite{Doi-Onuki-1992,Onuki-Book}. 
Note that eq.~(\ref{eq:eqm-04}) recovers the original two-fluid gel model for $\tau\rightarrow \infty$~\cite{Tanaka-JCP-1973,Levine-PRE-2001}, 
while for $\Gamma\rightarrow \infty$ it reduces to a linear viscoelastic Maxwell model~\cite{Fulford-Biorheology-1998,Bird-Book}. 
Decomposing the vector ${\bf f}^p$ into its longitudinal and transverse parts as
${\bf f}^p = {\bf f}^p_{\perp}+{\bf f}^p_{\parallel}$, where ${\bf f}^p_{\parallel} = \nabla^{-2}\nabla(\nabla\cdot{\bf f}^p)$,
and ${\bf f}^p_{\perp}=[{\bf 1}-\nabla^{-2}\nabla\nabla]\cdot{\bf f}^p$, we split eq.~(\ref{eq:eqm-04}) into two independent
equations. Transverse part of Stokes equation, (\ref{eq:eqm-02}), reads $0=\nabla^2{\bf v}_{\perp}+{\bf f}^p_{\perp}$.
It is easy to see that a zero shear viscosity of this medium is given by 
\begin{eqnarray}
 \eta_p &=& \eta+\tau\mu.
 \label{eq:eta_p}
\end{eqnarray}

\section{Perturbative analysis}
We introduce the stream functions, ${\bf v} = \left(\pa_z \psi, -\pa_x\psi\right)$ and ${\bf f}^p_{\perp} = \left(\pa_z\chi,-\pa_x\chi\right)$, 
where $\pa_x$ denotes the partial derivative with respect to $x$, so that they automatically satisfy the divergence-free condition.
The longitudinal force, ${\bf f}^p_{\parallel}$, is by definition a gradient of a scalar potential: 
${\bf f}^p_{\parallel} =-\nabla \phi$. 
The viscoelastic force is reconstructed from the scalar functions as 
$f^p_{x} = -\pa_x\phi+\pa_z\chi$ and $f^p_{z} = -\pa_z\phi-\pa_x\chi$.
The boundary conditions on the sheet surface read 
\begin{eqnarray}
 \nabla \psi (x, \eps b\sin(kx-\omega t),t) &=& \eps b\omega\cos(kx-\omega t) \hat{\bf e}_x,
 \label{eq:bc-psi}\\
 {\bf f}^p(x, \eps b\sin(kx-\omega t),t) &=&  {\bf 0}.
 \label{eq:bc-fp}
\end{eqnarray}
The first equation is the usual no-slip conditions for the velocity field ${\bf v}$. 
The second one requires that the network and the solvent must move together on the surface 
because no penetration of polymer chains across the sheet is allowed.

Due to the linearity of the governing equations, we replace time derivative with $-i\omega$ and obtain 
the equations for the functions $\psi, \chi$ and $\phi$:
\begin{equation}
 0 = \eta\nabla^2\nabla^2\psi+\nabla^2\chi,
 \label{eq:eqm-main-01}
\end{equation}
\begin{equation}
 0 = [\nabla^2-\kappa_{\perp}^2(\omega)]\nabla^2\chi,
 \quad
 0 = [\nabla^2-\kappa_{\parallel}^2(\omega)]\nabla^2\phi,
 \label{eq:eqm-main-02}
\end{equation}
where the inverse of dynamic screening length is given by
\begin{equation}  
 \kappa_{\perp}^2(\omega) = \frac{1+\nu_0-i\omega\tau}{\xi^2_{\perp}},
 \quad
 \mbox{and}
 \quad 
\kappa_{\parallel}^2(\omega) = \frac{1-i\omega\tau}{\xi^2_{\parallel}}.
\label{eq:kappa}
\end{equation}
The transverse and longitudinal viscoelastic lengths appearing in eq.~(\ref{eq:kappa}) are
\begin{equation}
 \xi_{\perp} = \sqrt{\frac{\mu\tau}{\Gamma}}
 \quad
 \mbox{and}
 \quad
 \xi_{\parallel} =\sqrt{\frac{(\lambda+2\mu)\tau}{\Gamma}},
 \label{eq:xis}
\end{equation}
where $\nu_0=\tau\mu/\eta=\eta_p/\eta-1$ is the rescaled solution viscosity.
Following the standard procedure, we expand $\psi$, $\chi$ and $\phi$ in powers of $\eps$ as
$\psi = \eps \psi_1+\eps^2\psi_2+\cdots$, $\chi = \eps \chi_1+\eps^2\chi_2+\cdots$
and $\phi = \eps \phi_1+\eps^2\phi_2+\cdots$.
The governing equations at each order keep the same form as Eqs.~(\ref{eq:eqm-main-01}) and (\ref{eq:eqm-main-02}),
and the boundary conditions at $O(\eps)$ order are
$\nabla \psi_1 = b\omega\cos(kx) \hat{\bf e}_x$, 
$f^p_{x1} = -\pa_x\phi_1+\pa_z\chi_1 =  0$, and
$f^p_{z1} = -\pa_z\phi_1-\pa_x\chi_1 = 0$ at $z=0$.
From the reflection symmetry in the problem: $\eps \rightarrow -\eps$~\cite{Taylor-Royal-1951,Childress-Book},
there is no net flow and all viscoelastic force vanish at infinity.
The first-order solutions take the following form:
\begin{equation}
 \chi_1 = [A_1e^{-\sqrt{k^2+\kappa^2_{\perp}(\omega)} z}+B_1e^{-kz}] e^{i(kx-\omega t)}+\mbox{c.c}, 
\end{equation}
\begin{equation}
 \phi_1 = [C_1e^{-\sqrt{k^2+\kappa^2_{\parallel}(\omega)} z}+D_1e^{-kz}] e^{i(kx-\omega t)}+\mbox{c.c},
\end{equation}
and
\begin{equation}
 \psi_1 = [E_1e^{-\sqrt{k^2+\kappa_{\perp}^2(\omega)} z}+(F_1z+G_1)e^{-kz}]e^{i(kx-\omega t)}+\mbox{c.c},
\end{equation}
where "c.c" stands for complex conjugate, and $A_1,B_1,C_1,D_1,E_1,F_1,G_1$ are complex numbers.
Six relationships between the integration constants, two from eq.~(\ref{eq:eqm-04}), and four from the boundary conditions, 
are found as
$A_1+\eta\kappa^2_{\perp}(\omega)E_1 = 0$, 
$2\mu\tau kF_1+(1-i\omega\tau)H_1 = 0$, 
$E_1+G_1 = ib\omega/(2k)$, 
$\sqrt{k^2+\kappa^2_{\perp}(\omega)} E_1-F_1+kG_1 = 0$, 
$\sqrt{k^2+\kappa^2_{\perp}(\omega)}A_1+kH_1+ikC_1 = 0$,
and
$kA_1+kH_1+i\sqrt{k^2+\kappa^2_{\parallel}(\omega)}C_1= 0$,
where $H_1=B_1+iD_1$. 
Solving these algebraic equations uniquely determines $\psi_1$ and thus the velocity field ${\bf v}_1$.


Next we proceed to determine the second order solution. 
Looking at eqs.~(\ref{eq:bc-psi}) and (\ref{eq:bc-fp}), the boundary conditions at $O(\eps^2)$ read
$\nabla\psi_2 = -b\sin(kx-\omega t)\nabla \partial_z\psi_1$ and
${\bf f}^p_2 = -b\sin(kx-\omega t)\partial_z {\bf f}^p_1$ at $z=0$. 
At this second order, a net flow at infinity is expected, which implies that time-independent terms should be added 
in the solutions for $\psi_2$, $\chi_2$ and $\phi_2$:
\begin{eqnarray}
\chi_2 &=& u_0e^{-\kappa_{\perp}(0)z} \nonumber \\
 	&+& [A_2e^{-\sqrt{4k^2+\kappa^2_{\perp}(\omega)} z}+B_2e^{-2kz}]e^{2i(kx-\omega t)}+
 	\mbox{c.c},
\label{eq:2nd-sol-chi}
\end{eqnarray}
\begin{eqnarray}
\phi_2 &=& v_0e^{-\kappa_{\parallel}(0)z} \nonumber \\
&+& [C_2e^{-\sqrt{4k^2+\kappa^2_{\parallel}(\omega)} z}+D_2e^{-2kz}]e^{2i(kx-\omega t)}+\mbox{c.c},
\label{eq:2nd-sol-phi}
\end{eqnarray}
and
\begin{eqnarray}
\psi_2 &=& w_0e^{-\kappa_{\perp}(0)z}+s_0z \nonumber \\
	&+& [E_2e^{-\sqrt{4k^2+\kappa_{\perp}^2(\omega)} z}+(F_2z+G_2)e^{-2kz}]e^{2i(kx-\omega t)} \nonumber \\
	&+& \mbox{c.c},
\label{eq:2nd-sol-psi}
\end{eqnarray} 
where $u_0, v_0, w_0$, $s_0$, 
$\kappa_{\perp}^2(0)=(1+\nu_0)/\xi^2_{\perp}$ and $\kappa_{\parallel}^2(0)=(1+\nu_0)/\xi^2_{\parallel}$
(see eq.~(\ref{eq:kappa})) are now real numbers.
(The constants $A_2$-$G_2$ are complex numbers, but are irrelevant to our aim below.)
The solutions, (\ref{eq:2nd-sol-chi})-(\ref{eq:2nd-sol-psi}), satisfy $\pa_z\chi_2 \rightarrow 0$ and $\pa_z\phi_2 \rightarrow  0$ at $z \rightarrow \infty$, 
which ensure that the polymer network and the solvent move together at infinity. 

The net pumping flow $u$ is the non-zero fluid velocity at infinity:
\begin{eqnarray}
 u &=& \eps^2\lim_{z\rightarrow\infty}\partial_z\psi_2 = \eps^2s_0, 
 \label{eq:u-def}
\end{eqnarray}
Thus the remaining task is to determine $s_0$ from the boundary conditions.
After some calculations, we arrive at a uniform hydrodynamic flow at infinity of (setting $\eps=1$)
\begin{eqnarray}
 u &=& \frac{b^2\omega k}{2}\left[1+\Delta'(k,\omega)\right],
 \label{eq:pumping}
\end{eqnarray}
where $u_N=b^2\omega k/2=(bk)^2c/2$ is the result for Newtonian fluids first obtained by Taylor~\cite{Taylor-Royal-1951}.
The viscoelastic correction $\Delta'$ is the real part of the function $\Delta$ given by
\begin{widetext} 
\begin{equation} 
 \Delta(k,\omega) = \frac{2\nu_0k\left[i(1+\nu_0)^{-1} (\omega\tau) \left(\sqrt{k^2+\kappa_{\perp}^2(\omega)}+k\right)-2k\right]}
{\xi_{\parallel}^2(\sqrt{k^2+\kappa^2_{\perp}(\omega)}\sqrt{k^2+\kappa^2_{\parallel}(\omega)}-k^2)(\sqrt{k^2+\kappa^2_{\perp}(\omega)}+k)(\sqrt{k^2+\kappa^2_{\parallel}(\omega)}+k)+2\nu_0k^2}.
\label{eq:delta}
\end{equation} 
\end{widetext} 
\begin{floatequation} 
\mbox{\textit{see eq.~\eqref{eq:delta}}} 
\end{floatequation} 

Equations (\ref{eq:pumping}) and (\ref{eq:delta}) constitute the main result in this letter.
In fig.~\ref{fig:u_vs_k}, $u/u_N$ is plotted as a function of $k\xi_{\perp}$ for varying $\omega\tau$, 
with the Poisson ratio of the network $\sigma=\frac{\lambda}{2(\mu+\lambda)}$ set 0.2.

\section{Results and Discussion}

\begin{figure}[htbp]
\begin{center}
 \includegraphics[width=0.99\linewidth]{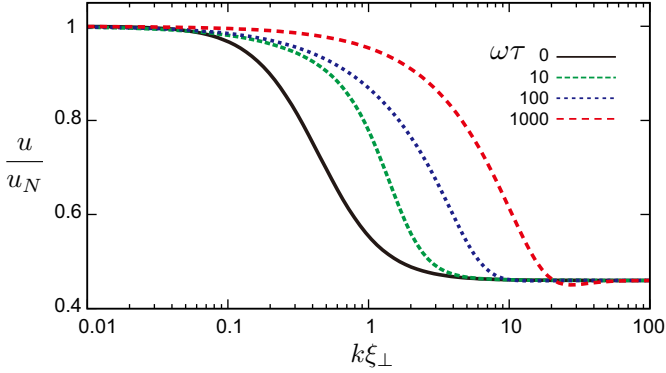}
\caption{The rescaled pumping speed, $u/u_N$, as a function of
the rescaled wavenumber $k\xi$ for different rescaled frequencies $\omega\tau$=0, 10,100, 1000.
The rescaled viscosity and the Poisson ratio of the network are set $\nu_0=\mu\tau/\eta=100$ and $\sigma=0.2$.}
\label{fig:u_vs_k}
\end{center}
\end{figure}

As seen in fig.~\ref{fig:u_vs_k}, the pumping flow is always smaller than that in a viscous Newtonian fluid.
The most important feature is that it strongly depends both on $(k,\omega)$ and on the viscoelastic parameters.
In particular, there are two qualitatively different regimes. 
In small wave number or short viscoelastic length regime, the viscoelastic force "diffusion" in eq.~(\ref{eq:eqm-04}),
$\Gamma^{-1}\mu\nabla^2{\bf f}^p$, becomes negligible compared to $\tau^{-1}{\bf f}^p$, 
and the model behaves like a linear viscoelastic Maxwell model.
Looking at fig.~\ref{fig:u_vs_k}, the pumping speed in this regime is unchanged from the viscous Newtonian case, {\it i.e.}, $u/u_{N} \rightarrow 1$ for all $\omega\tau$,
in agreement with the previous studies~\cite{Fulford-Biorheology-1998}.
In constrast, for large wave numbers, the stress propagation is dominated by the shear elasticity, and the whole medium acts mainly 
like a cross-linked gel. 
The diffusive term, $\Gamma^{-1}\mu\nabla^2{\bf f}^p$, now balances with $\mu\nabla^2{\bf v}$, 
leading to ${\bf f}^p\sim\Gamma{\bf v}$ (thereby directly suggesting $\dot{\bf u}\sim 0$).
The solvent flow through the virtually immobile rigid network is strongly damped via the friction
(just like a flow through a porous medium~\cite{Leshansky-PRE-2009}), 
and the far field hydrodynamic flow is accordingly decreased.
Taking $k\xi_{\perp}\rightarrow\infty$ limit in eq.~(\ref{eq:delta}), we obtain the asymptotic pumping speed
\begin{eqnarray}
 \frac{u_{\infty}}{u_N} &=& \frac{\lambda+\mu(1+2\eta/\eta_p)}{\lambda+3\mu}.
 \label{eq:u-limit}
\end{eqnarray}
This limit would mostly be relevant to dense viscoelastic cases, where $\eta/\eta_p \ll 1$ and $\xi_{\perp}\gg \xi_0$. 
In such cases,  we would obtain $u/u_N\sim (\lambda+\mu)/(\lambda+3\mu)$, 
which indicates that the pumping flow is ultimately determined by the elastic parameters only.

\begin{figure}[htbp]
\begin{center}
\includegraphics[width=0.95\linewidth]{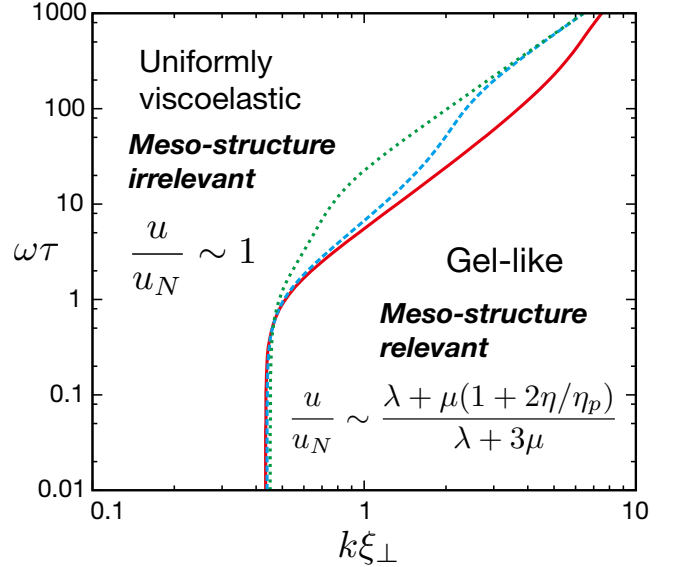}
\caption{The diagram of the pumping response on the $(k,\omega)$ plane. Two qualitatively different regimes are found.
The boundary line between the two regimes is determined as described in the main text, for different value of the solution
viscosity: $\nu_0=10$ (dotted, green), $\nu_0=100$ (broken, blue), and $\nu_0=1000$ (solid, red).}
\label{fig:diagram}
\end{center}
\end{figure}

Note that, while in fig.~\ref{fig:u_vs_k} we resaled the wave number $k$ with the viscoelastic length $\xi_{\perp}$,  
the characteristic length separating these two regime is not simply $\xi_{\perp}$, but is generally $\omega\tau$-dependent.
In fig.~\ref{fig:diagram}, we show the "diagram" in the $(k,\omega)$ plane.
Here the boundary line separating the two regimes is identified as the point at which $u$ decreases the half amount of
the full decrease in fig.~\ref{fig:u_vs_k}, that is, where $u=(1+u_{\infty})/2$.
From this diagram, we see that increasing $\omega$ (for a given $k$) enlarges the uniform viscoelastic regime, 
where the Newtonian-like response, $u/u_N\sim 1$, is obtained within the linear analysis.
This result for the first time clarifies in which condition the use of spatially uniform viscoelastic constitutive equations is justified.
Note that in this regime, nonlinear elastic effects have to be considered as shown in the previous studies~\cite{Lauga-PhysFluids-2007,Lauga-EPL-2009,
Fu-PRL-2007,Teran-PRL-2010}.

\begin{figure}
\begin{center}
 \includegraphics[width=0.99\linewidth]{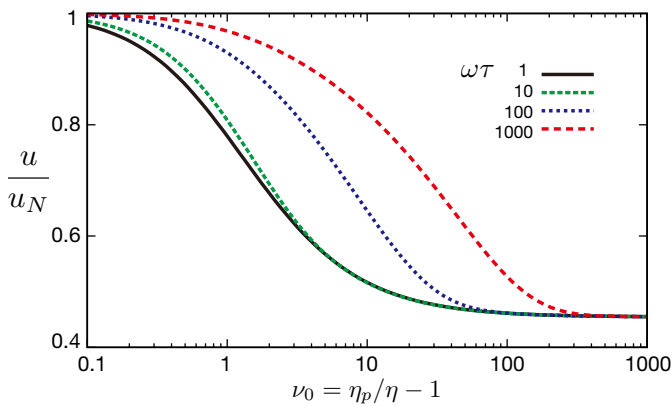}
\caption{The rescaled pumping speed, $u/u_N$, as a function of the rescaled viscosity, $\nu_0=\eta_p/\eta-1$,
for $k\xi_0=1$ (with the relation $k\xi_{\perp}=\sqrt{\nu_0}k\xi_0$), and for different $\omega\tau$ indicated on the figure.}
\label{fig:viscosity-dep}
\end{center}
\end{figure}

With the scaling of the friction constant, $\Gamma\sim \eta/\xi_0$, we obtain $k\xi_{\perp}\sim\sqrt{\nu_0}(k\xi_0)$.
The viscosity dependence for a fixed $(k,\omega)$ is thus known from $u/u_N=1+\Delta'(\sqrt{\nu_0}(k\xi_0),\omega\tau)$, 
which is plotted in fig.~\ref{fig:viscosity-dep} (assuming that $\tau$ is independent of $\nu_0$).
As a general tendency, the pumping speed descreases with $\nu_0=\eta_p/\eta$, in agreement with the previous theoretical and 
experimental observations~\cite{Lauga-PhysFluids-2007,Fu-PRL-2007}. 
For a very large solution viscosity, the viscoelastic length, $\xi \sim (\frac{\eta_p}{\eta})^{1/2} \xi_0$, also becomes very long, 
i.e., $k\xi_{\perp} \gg 1$, then the asymptotic value of $u/u_N$ is again given by eq.~(\ref{eq:u-limit}).
In stark contrast to the previous studies~\cite{Fulford-Biorheology-1998}, we predict the strong viscosity dependence of the pumping or swimming speed within the 
linearized viscoelastic model. 
We stress that this is a direct consequence of the mesoscale structure generic to entangled polymer solutions.

In the present two-dimensional geometry, the viscoelastic stress tensor can be given as
$\sigma_{xx}^p = -\phi +\partial_x\partial_z\theta$, 
$\sigma_{zz}^p = -\phi-\partial_x\partial_z\theta$, and
$\sigma_{xz}^p =\sigma_{zx}^p = -\frac{1}{2}(\partial_x^2-\partial_z^2)\theta$,
where the function $\theta(x,z)$ satisfies $\nabla^2\theta = 2\chi$.
The total stress tensor is the sum of the viscous and the elastic contributions: 
$\vecsg=-p{\bf 1}+\eta[\nabla{\bf v}+(\nabla {\bf v})^T]+\vecsg^p$.
Plugging the above expressions for $\vecsg^p$ into $\vecsg$, we obtain
\begin{eqnarray}
 \sg_{xx} &=& -(p+\phi) + \pa_x\pa_z\zeta, 
  \label{eq:sg-1}\\
 \sg_{zz} &=& -(p+\phi) - \pa_z\pa_z\zeta, 
  \label{eq:sg-2}\\
 \sg_{xz}=\sg_{zx} &=& -\frac{1}{2}(\pa_x^2-\pa_z^2)\zeta,
 \label{eq:sg-3}
\end{eqnarray}
where $\zeta=\theta+2\eta \psi$ satisfies
\begin{eqnarray}
 \nabla^2\nabla^2\zeta &=& 0.
 \label{eq:zeta}
\end{eqnarray}

The work done by the waving surface on a surrounding viscoelastic fluid per unit time (the rate of working),
$W$, is obtained by integrating the product of the velocity and the total force on the fluid over the surface
of the sheet $A$:
\begin{eqnarray}
 W &=& \int_A {\bf v}\cdot({\vecsg}\cdot\hat{\bf n}) \, dA.
 \label{eq:work}
\end{eqnarray}
Transforming the surface integral in eq.~(\ref{eq:work}) into the volume integral via the Gauss's theorem
and using the mechanical balance $\nabla\cdot\vecsg=0$ (obtained by adding eqs.~(\ref{eq:eqm-01}) and (\ref{eq:eqm-02})), 
we find $W = \int (\partial_kv_{i}) \sigma_{ik} dV$.
Plugging eqs.~(\ref{eq:sg-1})-(\ref{eq:sg-3}) into this, and using $v_x=\partial_z\psi$ and $v_z=-\partial_x\psi$,
we arrive at
\begin{eqnarray}
 W = S+\int dx \int dz \, \nabla^2\psi\nabla^2\zeta,
 \label{eq:W}  
\end{eqnarray}
where $S$ is the surface integral given by
\begin{eqnarray}
 S &=& \int dx (\partial_x\psi) \left(\partial_x-\partial_z\right)(\partial_z \zeta)|_{z=0}.
 \label{eq:S}
\end{eqnarray}
One can solve eq.~(\ref{eq:zeta}) if appropriate boundary conditions for the stress function $\zeta(x,z)$ are given depending on the problem at hand.
For example, for a freely swimming sheet, the total force acting on the sheet along the moving direction is zero:
\begin{eqnarray}
 \oint \hat{\bf e}_x\cdot{\vecsg}(x,z(x,t),t)\cdot{\bf n} &=& 0,
 \label{eq:force-free}
\end{eqnarray}
where $\hat{\bf n}$ is the unit normal to the surface, and the integral is over the surface of one period.
Once $\zeta(x,z)$ is known, the work $W$ is calculated according to eqs.~(\ref{eq:W}) and (\ref{eq:S}),
details of which will be published separately~\cite{Wada-unpublished}.

Finally, we briefly comment on the cross-linked gel (or solid network) limit,  $\tau=\infty$. 
This limit can be formally taken in eq.~(\ref{eq:delta}) without any mathematical singularity.
The resulting function behaves quite similar to eq.~(\ref{eq:delta}), with the characteristic length and time scale now replaced by $(\eta/\Gamma)^{1/2} \simeq \xi_0$
and $\eta/\mu$, respectively~\cite{Powers-PRE-2008}. 
Note, however, that cross-linked gels do not flow. In the gel limit, the no-slip conditions, eqs.~(\ref{eq:bc-psi}) (\ref{eq:bc-fp}), are inapplicable. 
While the no-slip boundary conditions for the network displacement ${\bf u}$ require to drag the whole solid material along the pumping direction, 
it is impossible because of progressively increasing restoring forces.
General slip boundary conditions, as well as convective nonlinearities, should be taken into account, as demonstrated for swimming in gels~\cite{Fu-preprint-2010}.

\section{Summary}
By analyzing the viscoelastic two-fluid model, we have studied fluid transport driven by a flexible sheet with a given traveling wave
deformation.
We used the linearized hydrodynamic model, but appropriately included non-local nature of the viscoelastic stress relaxations.
In contrast to a Newtonian fluid, the hydrodynamic pumping flow generated far from the waving boundary is strongly 
wave number and viscosity dependent. 
In particular, the two qualitatively different regimes of the pumping response are clarified. 
In one regime, the Newtonian-like response is obtained within the linear analysis, while in the other regime the pumping 
is substantially decreased due to the medium's gel-like character.
The boundary line separating the two regimes is shown on the plane spanned by the actuation wave number $k$ and frequency $\omega$.
The diagram illustrates when
the spatially uniform viscoelastic constitutive equations, such as a Stokes-Oldroyd-B model used in 
the previous studies~\cite{Lauga-PhysFluids-2007,Lauga-EPL-2009,Fu-PRL-2007,Teran-PRL-2010}, do provide good approximations. 

We conclude this paper with remarks on future extensions of this work.
It is important to explore effects of nonlinear convective terms omitted in our viscoelastic constitutive equation.
For finite amplitude undulations, this contribution may become substantial~\cite{Lauga-PhysFluids-2007,Fu-PRL-2007}.
More realistic geometries should also be considered for a detailed comparison with cilia and flagella beatings.
In particular, for free swimming, a crucial role of the undulating tail has recently been pointed out in ref.~\cite{Teran-PRL-2010}.
Dynamical shape changes responding to the hydrodynamic forces should also be allowed.
For those aims, numerical approaches would be more suitable.
Effects of meso-strucures in a viscoelastic medium upon the active pumping 
could be experimentally tested using artificial-cilia-mounted microfluidic devices developed
in refs.~\cite{Oosten-NatMat-2009,Vilfan-PNAS-2010}. 
Macro-scale experiments, similar to the one presented in ref.~\cite{Selverov-Phys.Fluids-2001}, would also be informative.

After we have submitted this paper for publication, we learned of the work by H. Fu, V. B. Shenoy, and T. R. Powers,
in which swimming in gels are studied using the two-fluid model~\cite{Fu-preprint-2010}.

\acknowledgments
We thank R. R. Netz and N. Yoshinaga for helpful discussions. 
Financial support from MEXT Japan (Grand No. 20740241) is acknowledged.
We are grateful to H. Fu and T. R. Powers for sending us their preprint and for the subsequent helpful discussions.

\end{document}